\documentclass[amsmath,amssymb,nofootinbib,prd]{revtex4}

\usepackage{setspace}
\usepackage{graphicx}

\usepackage{amsmath,amssymb,amsfonts,amsthm}

\usepackage{enumerate}

\begin{document}
\title{
Quantum panprotopsychism and a consciousness-centered universe}

\author{Rodolfo Gambini\footnote{Corresponding author. Email: rgambini@fisica.edu.uy}}
\affiliation{Instituto de F\'{\i}sica, Facultad de Ciencias, Igu\'a 4225, esq. Mataojo,
11400 Montevideo, Uruguay.}

\begin{abstract}
In previous publications, we have argued that a form of panprotopsychism based on quantum states and events offers a solution to the combination problem. This framework explains the emergence of complex phenomenal qualities and conscious subjects. Furthermore, the inherent openness of quantum mechanics allows consciousness --—and, more generally, phenomenal properties —-- to exert a causal influence. If the view proposed by quantum panprotopsychism is valid, it suggests that we inhabit a consciousness-centered universe. A world whose fundamental nature is phenomenal. This is at odds with the current view about the human condition that was strongly influenced by a science based on classical mechanicism, and led to nihilism and existentialism in late 19th-century Europe, and more recently to the rise of anti-foundationalists perspectives. The centrality of consciousness resulting from the incorporation of a quantum ontology into our worldview leads us to reconsider the nihilistic view and conclude that we live in a world in which a precise physical order leads to people capable of accessing a transcendent phenomenal realm.
\end{abstract}
\maketitle

\section{Introduction}

Physicalism provides a simple and unified perspective on the world, but it arguably falls short in offering a compelling explanation for the emergence of consciousness in humans and animals. Despite our direct awareness, conscious experience appears to remain inexplicable in physical terms. Physics can describe structures and processes, yet it fails to account for how subjective qualities arise from them.  Daniel Dennett \cite{dennett1} p.21 remarks that ``human consciousness is just about the last mystery to survive... Consciousness stands alone today as a topic that often leaves even the most sophisticated thinkers tongue-tied and confused.'' Consciousness is so fundamental to our existence that a life without it is deemed meaningless. Without consciousness, there is no personal life.

Panpsychism is the thesis that fundamental physical entities have elements like the mental ones. It posits that mentality is a fundamental and pervasive feature of the natural world. Rooted in the philosophical traditions of both the East and the West, this view has recently gained renewed attention in analytic philosophy. According to panpsychism, any object, described physically in third-person empirical terms, could also possess a phenomenal intrinsic nature. Typically, this perspective attributes mind-like qualities to elementary particles, such as quarks and photons. However, as quantum physics reveals, such attributions are misguided. The core concepts of quantum mechanics, as defined by its axioms, involve systems in specific states that produce events through interactions with other systems. This quantum ontology of states and events provides the foundation for a renewed form of panpsychism. We have analyzed this possibility in previous papers, noticing that when quantum systems are in entangled states, new properties and causal powers emerge that solve the different aspects of the combination problem \cite{c1,c2}.

An ontology grounded in classical physics necessitates a commitment to constitutive panprotopsychism, where all properties of a whole are functions of the properties of its parts. As Seager \cite{seagersep} notes, this perspective implies that ``the fact that I am conscious consists primarily in the fact that certain particles in my brain are arranged or interacting in a certain way.'' However, this form of panpsychism faces significant challenges, particularly with the combination problem. In contrast, a quantum ontology of states and events provides the foundation for a renewed form of panpsychism. This form of non-constitutive panpsychism posits the existence of both micro- and macro-experiences, without requiring that macro-experiences be fully grounded in micro-experiences. Although this view is incompatible with an ontology rooted in classical physics, it aligns naturally with that emerging from quantum physics, particularly for entangled states, as previously noted in \cite{lulegues, healey, teller}.

In our opinion, the most appropriate interpretation of the meaning of the scientific representation of the world comes from Van Fraassen's \cite{vfr2} scientific structuralism, influenced by thinkers such as Hertz, Mach, and Poincaré, and shared by many contemporary physicists such as Hawking. This considers physical descriptions as the result of processes of abstraction and idealization applied to an external reality. A reality that can only be referred to indexically, that is, by pointing to the object in question.
According to Van Fraassen, starting from real objects --—those that correspond to observable phenomena that are not intrinsically mathematical in nature—-- we assign mathematical properties to them through measurements. For example, the location of planets on the celestial sphere, as observed from Earth, illustrates this process. Using perspectives offered by phenomena, empirical descriptions are developed, such as Kepler’s laws, which are derived from meticulously gathered astronomical data. Theoretical work, then, involves the derivation of the empirical structures described in mathematical terms by Kepler’s laws, from a model of the Solar System and the laws of dynamics. The physical-mathematical description says nothing about the intrinsic nature of objects and is entirely compatible with a panpsychist view. However, it can provide information about the kind of objects to which we can assign phenomenal qualities and about the biological structures that can add consciousness to those qualities \cite{c1,c2}. 

In a recent paper \cite{c3} it was shown that self-consciousness, as the capacity to view oneself as a subject of experience, with the ability to interpret these experiences in conceptual terms and face the challenges that they pose, establishing goals and acting to deal with them, can be included naturally in a quantum panprotopsychic approach that solves simultaneously the different combination problems.  The causal openness of quantum systems provides self-conscious beings the ability to make independent choices and decisions, reflecting a sense of self-governance and autonomy. In this context, the issue of personal identity takes a new form free from the problems of the simple view or the reductive approaches \cite{korfmacher}. As shown in these works, if experimental evidence emerges of quantum phenomena in the brain that allows for the appearance of robust entangled states coupled to neuronal discharges, some form of Russellian-like panprotopsychism would be extremely plausible. This would require a deep revision of our contemporary conception of reality. It would be necessary to rethink the dominant worldview resulting from recent centuries of scientific activity incorporating the apparent centrality of conscious phenomena. The main objective of this work is to advance this task by assuming the quantum origin of consciousness.

In Section II, we recall the concept of quantum panprotopsychism and how it resolves the combination problem. It also provides a notion of subjectivity and self-awareness, as well as a characterization of persons, including the central role of agency. In Section III, we analyze how persons behave in our contemporary world, finding that their self-centered behavior is fundamentally flawed. In Section IV, we observe that, according to quantum panprotopsychism, conscious life is not a mere byproduct of the development of the universe, but its fundamental feature. We live in a consciousness-centered Universe. Finally, we identify some of the properties of what could be considered this final stage of consciousness that has manifested throughout time and civilizations, as well as the reasons for its exceptional nature. In Section V, we conclude with some remarks.

\section{Panprotopsichism, self-consciousness and persons}
 
Panprotopsychism holds that the intrinsic nature of the physical is experiential, making consciousness naturally explainable. However, when based on classical physics it struggles to account for how complex minds arise from interactions among micro-conscious entities. This issue, known as the combination problem, questions how ``small" conscious subjects with micro-experiences combine to form a unified ``big" conscious subject. Panpsychism views this as a fundamental challenge, as classical physics assumes that the properties of the whole are just functions of the properties of the parts. The classical physicalist approach attempts to explain consciousness via supervenience: a set of properties T supervenes on another set B just in case two things cannot differ with respect to the properties T without also differing with respect to their properties B. Two classical systems can only differ in their parts or in the relations among them. 
David Chalmers \cite{chalmerscp} has also categorized the combination problem into three sub-problems. The subject combination problem: How do individual micro-subjects form a unified conscious subject? The quality combination problem: How do different micro-experiences merge into a rich, unified qualitative experience? The structure combination problem, How do micro-consciousnesses organize into the complex structures we experience?

The subject combination problem \cite{c3} is especially important when examining self-consciousness and personal identity. It concerns how microscopic subjects of experience come together to form macroscopic ones like us. This involves explaining how micro subjects contribute to the emergence of a macro-subject. As we mentioned above, if one adopts an ontology based on classical physics, one needs a form of constitutive panprotopsychism, where each property of the whole is determined by the properties of its individual parts, suffering most acutely from the combination problem.

Non-constitutive panpsychism, in contrast, acknowledges both micro- and macro-experiences but maintains that macro-experiences are not grounded in micro-experiences. This form of panpsychism, as noted in \cite{lulegues, healey, teller}, aligns naturally with the ontology proposed by quantum physics. For example, Healey (\cite{healey}, Sec. 4) have observed that quantum mechanics satisfies  a ``Physical Property Holism,” which states that certain physical objects possess intrinsic qualitative properties and relations that do not supervene on the intrinsic properties and relations of their fundamental components. The emergence of novel properties in the quantum realm --—where events and properties are fundamental--— demonstrates the ontological novelty and non-separability inherent to quantum phenomena, particularly in entangled states \cite{c1}.

Entangled systems exhibit properties that cannot be fully explained by the attributes of their individual parts. A two-particle system in a singlet or triplet entangled state provides a striking example: Certain components of the system’s total spin are well defined even though the individual particles themselves lack well defined spin components. Such emergent properties may play a crucial role in the rise of consciousness within systems composed of multiple elements. Moreover, the number of emergent properties that do not derive from the characteristics of individual components increases exponentially with the number of constituents. For example, a system of n entangled spinning particles can exist in different pure states determined by $2^n$ independent coefficients, each state corresponding to distinct properties of the total system and different causal dispositions or powers to produce events in other systems \cite{lulegues, c1, davies3}.

The quantum analysis of states and events presented earlier offers a more promising approach to addressing the combination problem. Unlike classical physics, where the properties and states of a whole system are understood as the sum of its parts, quantum emergence challenges this notion. In entangled quantum systems, individual components lose their distinct identities, and only the properties of the whole remain well defined, as shown in the spin system examples. In other words, within a quantum system that maintains entanglement, such as one composed of entangled molecular systems \cite{fisher} or photons interacting with the neural system macro-level, qualities can emerge without preserving any trace of the micro-level qualities of the constituent entities. A similar phenomenon occurs in the structuring of perceptions and the emergence of subjects of experience \cite{c2,c3}.

\subsection{Proto subjects, consciousness and self-consciousness}

\subsubsection{Proto-subjects}

Whitehead’s approach to subjectivity is particularly well suited to quantum panprotopsychism, as it can be applied to fundamental physical processes involving interactions between proto-subjects and objects. According to him \citep{whiteheadaoi}, any physical system has a primordial form of experience with an emotional character. He states: \citep{whiteheadaoi}, p.176: ``the basic fact is the rise of an affective tone originating from things whose relevance is given... about the status of the provoker in the provoked occasion." This kind of subjective activity is referred to as prehension.

For Whitehead: \citep{WhiteheadProcess}, p.19 ``A prehension... is referent to an external world, and in this sense will be said to have a 'vector character'; it involves emotion, and purpose, and valuation, and causation," paralleling the way quantum states have dispositions to produce events. Precisely these volitive aspects are what could be associated with states that, in quantum mechanics, have been related with dispositions or tendencies to produce events. While he does not attribute consciousness to all entities in which prehensions or events occur, he maintains that the occurrence of events in physical systems is always preceded by the prehension of a proto-subject.

The proto manifestations of prehensions in simple quantum systems extend from the system’s preparation in a particular state to its measurement. As discussed in \citep{c2}, preparation places the system in a defined state, often linked to the measurement of a prior system, resulting in an observable event. Generically, when a system in a certain state is measured, a random choice with probabilities given by the Born rule determines the resulting event, and this measurement either alters or annihilates the previous state.

Schematically, during preparations, events put the subject in a specific state, while in measurement, the subject in a particular state generates an event in an object at the conclusion of a phenomenal process. In the first case, a prehension arises with the event that defines its properties, especially its phenomenal qualities. In the second case, it ceases with the production of an event through interaction with other physical systems. The first process produces a phenomenal quality or proto perception. The second represents the culmination of the phenomenal process through an action, a proto-volitional act linked to the selection of an event.

Quantum states manifest themselves through the disposition to produce events during measurements. Each event is preceded by the prehension of a proto subject of experience. A key moment in this process --—the critical point--— is the selection of a specific outcome during measurement. When fundamental physical properties, such as the position of an electron, are measured, these selections occur randomly. Prehensions represent the phenomenal counterparts of the processes that precede measurements or follow preparations. In these processes, the relevant state where the prehension occurs is the state of the micro-system in interaction with the state of the measuring or preparation device \cite{peresspri}.

This perspective is agnostic with respect to the status of quantum states. In our previous papers, we adopted an interpretation of states as dispositions to produce events. This view is consistent both with an ontic or epistemic notion of states. In the ontic case, the state describes the disposition of the quantum system in a given state. In the epistemic case it describes the disposition that the observer attributes to a system whose phenomenal properties could transcend the physical description in order to describe its future behavior. At no point would the arguments about the combination problem or the personal identity problem\cite{c1,c2,c3} cease to be valid if we adopted ontic or epistemic positions consistent with this view about the dispositional nature of states. Epistemic views are closer to the initial Copenhagen interpretation, championed by Bohr, Heisenberg, Pauli, and others, who regarded the wave function as a "probability wave" and were agnostic about the need for a more fundamental reality to support it. This and other epistemic interpretations \cite{ep1,ep2} consider that the state is a representation of knowledge, information, or belief \cite{leifer} about the dispositions of a system to produce events, and that the collapse of the wavefunction can be explained as the effect of acquiring new information. As we mentioned, following Russell and van Fraassen, physics provides mathematically precise descriptions of processes and structures. Quantum theory, expressed in terms of systems in certain states that produce events, summarizes the rigorously predictable properties of a reality that could transcend mathematical description and, in quantum panprotopsychism, is assumed to possess a phenomenal nature. One says that a system in a certain state has a property $a$  if when one measures the observable associated with this property, one observes with certainty that the observable takes the value $a$. For example, a spinning particle has a component along a direction $z$ given by $z,up$ if each time I measure its $z$ component i obtain the result $up$.

The essential components of a conscious volitional act in higher conscious beings should parallel those previously discussed: an object --—the brain—-- and a subject’s prehension, described by the state of a specific set of neurons linked to an entangled quantum system. For instance, in Fisher’s model \citep{fisher}, this could involve a multi-entangled network of Posner molecules, where a prehension takes place, triggering an event, namely, the firing of specific neurons that initiate an action. In essence, a subject's prehension corresponds to the phenomenal aspect of a system in a given state, with its duration spanning the interval between the event that generates it and the event it subsequently produces in another system.

In this context, the issue of subject combination is easily resolved. When two proto subjects interact and become entangled, they merge into a single entity, losing their individual identities in the process. As a result, the new subject corresponds to the total system in its entangled state. Drawing an analogy between elementary quantum measurements (or preparations) and mental processes, the quantum system can be understood as a multi-particle entangled state. Some components of this system interact with the neural system that either changes the quantum state or is changed by it. In the case of a preparation, this interaction manifests itself as a new perception, whereas in a measurement, it appears as an act of volition.

As discussed previously, when a quantum system is fully entangled, the individual identities of its components are lost. Consequently, the only relevant subject is that associated with the entire entangled system. This realization has been instrumental in addressing the quality and grain combination problems. Once the total entangled system is acknowledged as the subject of successive experiences, the subject-summing issue can be resolved in the same manner as the other combination problems. \citep{c1,c2}.

\subsubsection{Perceptions in conscious beings}

Perceptions have informational content. They give us the only (immediate) insight concerning reality, something which can never be obtained from conceptual, or logical, means. As we have seen, perceptions follow events that play the role of preparations of the state. Our perceptions are rich not only in qualities but also in structure. The wealth of our perceptions and their organization in complex spatial or temporal representations that combine data from different senses seem to be very different from the structure of the neural organization. A particularly pressing aspect of this problem, called the "granularity problem", is the concern that experiences appear fluid and continuous in a way that is at odds with the discrete and particularized structure of the neurological properties of the brain. Chalmers poses the problem as follows: \cite{chalmers} p. 5: ``Macrophysical structure (in the brain, say) seems entirely different from the macrophenomenal structure we experience. Microexperiences presumably have structure closely corresponding to microphysical structure..., and we might expect a combination of them to yield something akin to macrophysical structure. How do these combine to produce a macrophenomenal structure instead?'' The problem of granularity would not arise in our model since the entangled quantum system would only have, as we saw, global properties \cite{c2}. There is, however, a second aspect of the problem of structure: our macro experiences have a rich structure, which involves the complex spatial organization of the visual, auditory and the other senses. How can micro-experiences lead to the rich structure of the perceptions and its strong informative content? Phenomenologically \cite{husserl,c1}: I am sitting in the room in front of the computer and I am aware of the furniture, books, lights, etc. I am also aware of the space behind me and the sounds coming from a nearby construction, the blue sky I see through the window, the current morning moment, the recent breakfast, and the day's tasks that I have to perform. The modes of manifestation of consciousness should be understood not as psychic occurrences similar to physical ones, but as meaningful experiential engagements in their own right. These take place in the context of a world of intentional implications and motivations. If this informativeness of perceptions developed due to Darwinian evolution, some form of efficacy of the phenomenal contents is needed to change the adaptive capacity. In recent decades, the conviction has increased that consciousness is present not only in mammals and birds, but also extends to primitive animals such as worms or jellyfish \cite{gija}, and there are multiple examples of the evolution of our perceptive abilities\cite{low}.

Animal behavior is entirely driven by information derived from their perceptions, without any form of conceptual analysis. As we move up the evolutionary scale, these perceptions become increasingly complex and informative. From a Darwinian perspective, this implies that the phenomenal aspects of perception must contribute to adaptive advantages, enabling natural selection to act upon them. Within the framework of classical physics, such physical-phenomenal interactions would be impossible because of its causal closure. However, interactions between the nervous system and its phenomenal counterpart would be essential. The causal openness of quantum mechanics allows such interactions to occur. In a quantum system such as the one described, the phenomenal could influence the physical by determining which neuronal firing events occur during the interaction between the quantum state and the brain's neural network, which would act as a measuring device. Better structuring of perceptions will lead to better object identifications, and consequently, to a greater capacity for adaptation.

Animals and young infants exhibit a direct, non-reflexive form of consciousness. They have consciousness but not self-consciousness \cite{c3,lrb}. Wittgenstein in his Tractatus \cite{wittgenstein} uses the following metaphor: consciousness is for the infant ``like that of the eye and the field of sight. {\it But you do not really see the eye.} And from nothing {\it in the field of sight} can it be concluded that it is seen from an eye”.  And as Choifer \cite{choifer} adds p. 339: ``For the one who is at the origin [with direct inner acquaintance] there is no understanding of him/herself as a subject (of one’s own mental states), i.e., he/she is not yet a reflectively conscious subject. The one who is at the origin is not in the world but one with the world. The ‘origin’ perspective constitutes a unique, subjective first-hand encounter with a phenomenon. However, the price for this uniqueness is that one cannot, without stepping outside (i.e., switching to the ‘outside’ perspective) compare or share one’s unique and private point of view with anyone else’s. In this sense the point of view from origio is mute.” It is a perspective without any conceptual content \cite{persons}. However, most animals possess not only sensitivity –—they are sentient–— but also the ability to perceive, meaning that they can discern whether the world is in a certain way or another. This capacity for perception is crucial to the lifestyles these animals adopt, enabling them to recognize potential prey or threats, for example.

There appear to be no clear boundaries beyond which direct inner awareness and certain cognitive-like processes cease to exist. Nervous systems are present not only in vertebrates, but also in mollusks, arthropods, and even worms. Ultimately, varying levels of sensitivity and goal-directed behavior can be observed across all living organisms. The emergence of consciousness is a gradual process that begins in the early stages of animal evolution. If we accept that some degree of subjective experience exists even at primitive developmental stages, we should consider that a complete ontology of the physical world should account for some form of phenomenal experience in non-living entities, as we have suggested above.

\subsubsection{Self-consciousness}

Self-awareness entails the ability to focus on oneself and consciously recognize one's attitudes and dispositions. It involves awareness of perceptions, emotions, feelings, and behaviors, as well as recognition of one's own body and its relationship to others. To be self-conscious is to be aware that one is a conscious subject, which implies not only recognizing one's conscious experiences but also being certain that they occur within oneself, the one who acts and experiences through this body.

Animals already possess the ability to distinguish their own bodies from other objects, a capability derived from their perceptions. Similarly, they recognize objects such as stones or balls based on their effects and behaviors \citep{Gazzaniga}. The structured nature of perception likely has an evolutionary basis, as discussed previously \citep{c2}, while conceptual analysis is rooted in language. Animals appear to exhibit preconceptual abilities; for example, they can correct mistakes through perception, which in some cases requires distinguishing between correct and incorrect outcomes. They achieve this despite lacking language or a formal concept of truth and falsity. However, while animals can classify objects by recognizing relationships and analogies ---for example, differentiating bananas and pineapples from a piece of wood--- this does not necessarily mean that they possess a conceptual understanding of food \citep{dieguez}.

Neuroscientists have identified that self-consciousness is primarily located in regions of the left hemisphere of the brain closely linked to language centers. This has been confirmed through studies on split-brain patients, where the corpus callosum, the bridge between the two hemispheres, is severed, resulting in two disconnected brain halves. Despite this, patients report feeling completely normal and not noticing any changes after surgery. However, when a light is flashed independently to each hemisphere, an interesting phenomenon occurs: if the light is presented to the left hemisphere, the patient reports seeing it. In contrast, when flashed to the right hemisphere, the patient verbally denies seeing anything. However, if given the opportunity to respond using a Morse code key, they successfully press it with their left hand \cite{c3}. This suggests the existence of some kind of parallel processing with responses to stimuli that are not recognized by the self-consciousness center.

The left hemisphere houses the conceptualization center, referred to as ``the interpreter" by \cite{Gazzaniga}. In Husserlian terms, it assigns a noema to each intentional perception. This interpreter module constructs explanations based on the information it receives. It appears to be uniquely human and specialized in the left hemisphere. Its continuous drive to form hypotheses serves as the foundation for human beliefs, which in turn shape our cognitive processes. ``Our subjective awareness arises out of our dominant left hemisphere's unrelenting quest to explain these bits and pieces that have popped into consciousness" \cite{Gazzaniga}, p. 102. Consequently, neurophysiological evidence suggests that multiple phenomenal processes can coexist within the brain, although we are only consciously aware of some of them. As Dennett \cite{dennett1} points out, while our brains can be placed on a continuum with insect nervous systems at the simplest end and dogs, dolphins, and chimpanzees closer to us in complexity, this perspective overlooks a crucial fact: humans are the only species that ask questions. This trait is directly linked to the interpretational modulus of the brain, as previously mentioned, which is uniquely human and absent in other animals. Although evolutionary processes suggest continuity, they do not negate the profound qualitative transformations that arise in humans and are at the basis of the advancement of human knowledge and culture. Indeed, these transformations encompass not only the concepts conveyed by language but also any type of symbolic expression.

An animal exists within an environment to which it is fully adapted under favorable conditions, forming a closed and purposeful system with its surroundings. In contrast, human beings do not merely inhabit an environment. They actively construct it, shaping their surroundings through conceptual activity \cite{cassirer2}. This ability allows humans to break free from the constraints of a predetermined way of life. As subjects of experience, individuals perceive, act, and express themselves conceptually and symbolically, driven by their capacity for autonomous action. They develop values, beliefs, thoughts, behaviors, choices, and decisions. The self is the bearer of intentional states, capable of reflecting on them and acting with freedom and responsibility. It evaluates its goals and beliefs, contemplates its future and objectives, and chooses its course of action or way of thinking. 

Only persons are aware that they are subjects of experience. There is a fundamental difference between merely experiencing needs or fears --—such as pursuing prey or fleeing —-- and possessing a first-person perspective. This perspective is unique to persons and constitutes their defining characteristic. It requires conceptual elaboration: the ability to conceive of oneself as the center of action and experience. While an animal recognizes its body through perception, expressing needs, perceptions, desires, or beliefs necessitates conceptual capacity. The pronoun "I" serves to denote the entity capable of perceiving itself as the subject of perception, decision-making, and volition. It is not merely about being a subject, but about experiencing oneself as a subject. The first-person perspective is what enables individuals to be conscious of their intentional states, a capacity exclusive to persons \citep{lrb}.

While non-human animals possess only biologically predetermined goals that they cannot alter, human persons have the capacity to evaluate and modify their objectives. This strongly suggests that self-consciousness, self-determination, and free will are deeply interconnected in persons. If Samuel Alexander’s maxim, “To be real is to have causal powers” \citep{alexander}, holds true, then there must be a real agent at the origin of the choices that shape our reflective processes. The nature of this agent would ultimately be defined by its free actions. The continuity of personal identity does not reside in the self as it exists at any given moment but in the succession of acts that respond to our goals, existential concerns, and moral responsibilities. Thus, a person cannot be reduced to a single act of perception or a mere sequence of such acts. Instead, persons are fundamentally constituted over time through conceptual elaboration and successive choices to act \cite{c3}.

Persons have the capacity to evaluate and modify their goals. Freedom is not mere randomness in action, but the ability to choose among different courses of conduct, guided by various impulses or orientations, often shaped by culturally established beliefs, values, and ideals. Every act of thought, composition, painting, helping others, or even choosing inaction involves a decision. At any moment, we could act differently and we are constantly challenged to act better. The self experiences dissatisfaction with given answers and continues searching until it finds alternatives that quiet its unrest. Every act of creation is an act of choice. Not only freedom is the ontological condition of ethics as noted by Foucault, but also of aesthetics and all forms of symbolic activity \citep{foucault}. If quantum indeterminacy applies to our mental processes, then personal freedom may involve more than mere randomness \citep{persons}. However, this would require identifying an ethical and aesthetic source of normativity or orientation accessible to consciousness. It is clear that we can not only conceive of a range of possible aesthetic and moral options, but also assess their relevance to specific situations with a reasonable degree of accuracy. This capacity for choice often seems to originate in intuition, functioning more as perception and evaluation, as Whitehead suggests, than as rational analysis.

\subsection{Person’s characterization and personal identity}

Strawson differentiates between two aspects of the self: the "I," which represents the subject of experience, and the "me," which constitutes the objective aspect formed by our self-image. This distinction, originally proposed by William James, separates the self as the knowing subject from the self as ``an empirical aggregate of objectively known things" \citep{persons}. James associates the ``me" with the self-image we construct or seek to project, suggesting that individuals can develop multiple ``me-selves" depending on their social roles.

Scholars such as Tagini \citep{tara} argue that the "I" corresponds to self-referential awareness linked to phenomenal consciousness rather than implicit mental representations. In contrast, self-awareness related to the "me" emerges from conceptual construction and reflective cognition. According to James, the "I" is the conscious subject at any given moment, actively perceiving interpreting and acting, whereas the "me" encompasses the explicit sense of self, shaped by distinct, identifiable attributes that contribute to one's self-concept \citep{tara}. {\em Thus, the "me-self" operates as a cognitive construct, effectively forming a "theory" of the self} \cite{epstein}. This construct is susceptible to error and misinterpretation; it offers a first-person perspective on how our brain organizes perceptions and actions initiated by the "I." Both the "I" and the "me" evolve over time along with the brain.

In a previous paper \cite{c3}, we identified three levels involved in the processes of perception and reflection, which may ultimately lead to an action of the ‘I’. This framework provides a notion of personal identity that aligns with the form of quantum panpsychism proposed here. The key elements that define a self-conscious person at a given moment are \cite{c3}:
i) The information embedded within the neurological structure of a brain, which is capable of exchanging signals with sensory organs and coordinating interactions with the body.
ii) The prehensions occurring at this moment.
iii) When prehensions lead to volitional or conceptual decisions: the agent who selects among various possible options related to the subject at hand. Through this selection, the agent contributes to shaping the concepts and beliefs encoded in the brain, which, in turn, influences other bodies through its interaction with the physical world.

In a framework of quantum panprotopsychism --assuming, for instance, that is implemented by the model of cognition based on Fisher's approach ---- each act would correspond to an interaction between Posner molecules’ entangled quantum systems and a network of neuronal synapses, manifesting as neural discharge events. As a result, every action would have a well-defined physical counterpart. However, due to quantum indeterminacy, two identical brains experiencing the same prehension could exhibit different behaviors depending on the specific choice arising from the quantum system’s dispositional state.

Note that the ``me" need not be included as an independent element. It is a product of memory, prehensions and actions that gave form to the brain, but also of psychological and physical conditioning that limit our free decisions and plays a determining role in the resulting image, for example, having bad memory, being shy or irascible, clumsy, or gifted at sports. These conditioning factors in themselves are not part of the ``me", but the way we assume them contributes to the image we construct of ourselves.  

As mentioned above, both animals and humans possess consciousness, but only humans develop a sense of self. This distinct human trait is closely linked to their conceptual abilities and their capacity to engage with highly structured symbolic systems, such as language or art. As Pickering \citep{pickering} (p. 9) suggests, the self ``participates in the continuous creative advance that, by bringing the past into the present, opens up a space of possibilities from which choice serves as a bridge to the future." The essence of the self lies in these choices, yet they do not arise in isolation. Rather, they are shaped and influenced by the brain’s contents --—our memories, knowledge, and beliefs—-- as well as by our perceptions, the range of possibilities available in our phenomenal experiences, and the will to act.

The three levels that intervene in the process of reflection followed by a free action of the "I” are involved in the personal manifestations, but actions resulting from conceptual elaboration are the main cause of the development of our brain capacities and of its activities, and therefore the defining elements of the self. In a previous paper \cite{c3} we have shown that the characterization of personal identity proposed there, given by the three conditions mentioned above, is free from the criticisms and paradoxes noticed in other approaches. As Cassirer observes \cite{cassirer2} in his study of the development of the self in mythical thought p.157: ``It is not mere meditation but action which constitutes the center from which man undertakes the spiritual organization of reality. It is here that a separation begins to take place between the spheres of the objective and subjective, between the world of the I and the world of things. The farther the consciousness of action progresses, the more sharply this division is expressed, the more clearly the limits between I and not-I are drawn. Accordingly, the world of mythical ideas, precisely in its first and most immediate forms, appears closely bound up with the world of efficacy."

\section{The modern personal condition: from enlightenment to antifoundationalism}

The Enlightenment has its roots in the scientific revolution of the 16th and 17th centuries. For Kant, enlightenment signifies the commitment to think independently; to use one's own reasoning to decide what to believe and how to act. Enlightenment thinkers generally held strong confidence in the human intellect, believing in its ability to attain systematic knowledge or at least well-founded beliefs about nature and to grasp their broader philosophical implications. This intellectual optimism coexisted often with a persistent tension with organized religion \cite{SEPenl}.
Enlightenment in its most radical form was born from the conviction that any tradition, argument, or belief that conflicts with the laws of nature expressed in mechanistic, mathematically verifiable terms must be rejected. It was Spinoza who first established, based on his doctrine of the existence of a single substance, that "Nothing then can happen in Nature to contravene her own universal laws, not anything that is not in agreement with these laws or that does not follow from them" \cite{ttp} p. 126. He assumes that the same laws of motion and cause and effect apply in every context and everywhere dismissing the 'supernatural' as a total figment of our imagination \cite{Israel}. With the enlightenment, science became a fundamental part of philosophical reflection and its most solid foundation.
Although the Enlightenment was marked by strong confidence in human reason, the period is equally defined by the emergence of empiricism both in scientific practice and in theories of knowledge. While thinkers like Spinoza and the rationalists placed unwavering trust in reason, enlightenment enthusiasm was less about reason as an isolated source of knowledge and more about the broader capabilities of human cognition. Newton demonstrated that natural science could succeed without relying on a priori self-evident first principles. The Enlightenment’s characteristic skepticism toward all claims of obscure or unquestioned authority --—particularly those of religion—-- extended to metaphysics as well \cite{SEPenl}. The prevailing attitude of the era --—marked by a distrust of authority and confidence in one’s own judgment—-- embodies the Enlightenment ideals of individualism and self-determination.

Hume was a staunch admirer of Newton as a scientist and took Newtonian physics as a model for how to build his treatise about human nature. However, Hume's skepticism questions the assumptions of scientific inquiry with the problem of induction. In short, it is possible for events that have occurred regularly according to the laws of nature to cease to obey these laws. It is possible for gravitational interaction to become repulsive, or for the moon to start showing its other side, which contradicts our past experiences, but not any logical principle. The fact that the laws of nature that we assume to be valid need not always hold was anticipated by Newton. In Principia, Book III, Rule IV, Newton's fourth rule of reasoning states: "In experimental philosophy, propositions obtained from phenomena by induction are to be considered exact or very nearly so in spite of all contrary hypotheses, until other phenomena render such propositions more exact or subject to exceptions." Consequently, we must consider a rule as valid until new evidence leads us to refine or replace it. But Newton is establishing a methodological principle to construct laws based on experience, admitting the limitations that all experimental knowledge has \cite{persons}. 

Paradoxically, the determinism of the new physics ultimately leads to the ideal of societies inspired by a notion of autonomy that allows for questioning the very laws that regulate them. This optimistic view of the world seems to arise from an incomplete analysis of the consequences of the mechanistic vision, which only reached its full potential in the 19th and 20th centuries. In fact, in the last centuries this legacy contributed to the rise of secular, nihilistic attitudes that consider consciousness as an illusion or an epiphenomenon without any causal efficacy, and freedom as a fiction. As Bristow puts it \cite{SEPenl} “Instead of being represented as occupying a privileged place in nature, as made in the image of God, humanity is represented typically in the Enlightenment as a fully natural creature, devoid of free will, of an immortal soul, and of a non-natural faculty of intelligence or reason.” If we adopt the quantum ontology that underpins the panprotopsychism considered here, a different view of our place in nature and its meaning emerges. To explore what this shift might entail, we must first examine the views mentioned above on the human condition.

\subsection{Nihilism}

As physics advanced and the world came to be understood as a mechanistic system governed by classical laws, various intellectual movements, ranging from nihilism to existentialism, emerged in late nineteenth-century Europe. This shift coincided with the decline of the pre-modern religious worldview, giving way to an increasingly secular and scientific perspective. Humans who formerly believed they were children of God and lived at the center of the Universe began to lose all special status, first with Copernicus and then with Darwin, becoming another animal. The mechanistic vision of the world gradually led to the denial of any transcendent moral framework and contributed to the development of the defining experiences of modernity: anxiety, alienation, boredom, and meaninglessness \cite{sepexist}. The term nihilism gained prominence in 1862 through Ivan Turgenev’s novel {\em Fathers and Sons}, in which he used it to characterize the scientistic worldview of his protagonist Bazarov, who advocates a doctrine of complete negation \cite{nishitani}.

To be a person is to be dissatisfied with one's condition, to strive to achieve a superior or simply different way of being. As Kaufmann puts it, “Man is the ape that wants to be a God” \cite{kaufmann}.  A being that exists as an endless project inevitably collides with the reality of death --—the denial of all further becoming. Thus, the need for transcendence arises precisely from an inherent lack of fulfillment. We strive to shape ourselves through our choices and actions, trying to conform to what will supposedly make us better, happier, or more powerful. As a historical process, this view of the human condition gradually developed following the growth of modern natural science and technology that treat us and the world as orderly resources for technical application \cite{Heidegger2} and became dominant with the industrial revolution and the rapid development of science in the XIX century, in particular Darwinism. 
Nietzsche summarizes the historical situation in his famous statement 'God is dead', first uttered in The Gay Science \cite{(1882)}, referring not only to the Christian deity, but to any belief that the world was governed by a set of absolutes ruling in a supra-sensory world of abstract definitions of reality, such as that originating in ancient Greece with Plato \cite{nickbea}.

The feeling of incompleteness and finitude seems to be the default characteristic of people in many societies. However, in the West, the acute process of secularization that accompanied the advances of scientific rationality has made the awareness of this human condition particularly severe, and the technological growth has led developed societies to enter in a new stage that Nietzsche denounced as the appearance of the last man living in Bauman's liquid modernity.
He summarizes \cite{Bauman} p.28 the evolution of the last centuries as follows. "As Lessing pointed out a long time ago, at the threshold of the modern era we have been emancipated from belief in the act of creation, revelation, and eternal condemnation. With such beliefs out of the way, we humans found ourselves 'on our own' - which
means that from then on we knew of no limits to improvement and
self-improvement other than the shortcomings of our own inherited or acquired gifts, resourcefulness, nerve, will and determination. And whatever is man-made, men can un-make. Being modern came to mean, as it does today, being unable to stop and even less able to stand still. We move and are bound to keep moving not so much because of the 'delay of gratification', as Max Weber suggested, as because of the impossibility of ever being gratified."

We are witnessing, at the beginning of the third millennium, the culmination of the realization of the last man. This figure embodies the very aspirations of modern society and Western civilization, whether consciously or unconsciously.  The last man embodies a society devoid of ambition, passion, or the desire to fully engage with life. He sees no intrinsic meaning or value in existence and passively accepts this emptiness without seeking alternatives. He avoids self-transcendence and refuses to challenge his limitations. Such a figure can only emerge from a humanity that has become apathetic. A society that loses its ability to dream, strive, and embrace collective risks.  In our time, we witness a society that appears unable to confront the urgent threats of its own future, choosing instead to ignore or deny their very existence. The pursuit of economic growth and technological dominance is carried out without much regard for the consequences for both our planet and future generations.

We have not yet assimilated the implications of the quantum description of the universe and what it entails in overcoming the mechanistic vision resulting from classical physics. Just as Newtonian mechanism underpinned the development of modernity, a panpsychist interpretation of quantum mechanics could open doors to a more respectful approach to nature and human existence. In the context of the quantum panprotopsychic personal characterization \cite{c3}, freedom is what sets human existence apart from that of other beings. We are free and accountable for who we are and what we do. We are always beyond or more than ourselves due to our ability to interpret and assign meaning to whatever constrains or defines us \cite{sepexist}. In contrast, for the last man, freedom is largely reduced to a series of trivial choices. Its exercise of freedom amounts to the indefinite postponement of any truly significant decision. We have achieved excessive power over the world at a time when we were least prepared for it, because we have lost the ability to direct our actions by prioritizing respect for people or nature, which we have transformed into a mere resource. For Nietzsche, the last man, with all his technological capacity, is not capable of assuming the responsibilities that his time demands of him. He is a man who has lost his moral compass. In the prologue to his Zarathustra, Nietzsche announces that the time of the most despicable man is coming, who is no longer able to despise himself. A dehumanized conformist, alienated, indifferent and baffled [person], directing psychological energy into hedonistic narcissism or into a deep resentment that often explodes in violence. 

Does Nietzsche provide any hint as to how to overcome this easily recognizable state in the last man? Heidegger \cite{thinking} acknowledges a possible answer in part two of Thus Spoke Zarathustra, where he says of the last man: “For that man be delivered from revenge: that is the bridge for the highest hope for me, and a rainbow after long storms.” and he makes explicit that by revenge he means “the will revulsion against time and its ‘It was’”, that is, the revulsion towards the transitory and passing nature to which our way of experiencing time has been reduced. Nietzsche attempted to overcome this condition by affirming the eternal return of the same. Let us live with joy, loving the moment as if it were repeated infinitely in a temporality that flows and returns like a circle. Let us embrace the death of God with joy and laughter, with a profound sense of liberation.  Without judging whether there is an alternative way of living and experiencing the passage of time, it seems clear that for the last man who permanently desires, fears, strives, or simply seeks distraction, there cannot be an affirmation of the present.

Socially and historically, this view is directly related with the interpretation of the physical sciences in purely mathematical terms inspired by Galileo and the mechanistic understanding resulting from classical physics. It remains valid today, even though the physical advances resulting from the development of quantum mechanics and the philosophical analysis of the last two centuries since the time of Hertz and Poincaré have revealed its falsity, as we mentioned above. \cite{vfr2}. Husserl observed that for Galileo, the mathematical models describe a reality which is mathematical in nature. Consequently, as he emphasized in Crisis \cite{crisis2}, mathematizability becomes a criterion for existence, and there is nothing beyond what is mathematically described. But this physicalist view is simply unfounded.

Papineau \cite{papineau} defines physicalism as the position that everything that exists is physically constituted. That is, all entities are either physical or depend on physical properties: if two entities can influence a physical system differently, then they must differ in their physical components. According to Papineau, the rise of physicalism stems from the belief in the completeness of physics, specifically, the principle of causal closure, which holds that every physical effect has a physical cause. He acknowledges, however, that in the context of quantum mechanics, this principle must be reformulated: rather than strict causality, the theory stipulates that the probabilities of physical events are determined by prior physical conditions. Despite this, Papineau downplays the relevance of this reinterpretation and proceeds as if the classical causal closure still holds. This dismissal raises concerns about the consistency of his physicalism. In quantum mechanics, events unfold through the realization of one among several probabilistic outcomes, with no underlying physical cause dictating which occurs. For example, the appearance of a dot on the photographic plate in the double-slit experiment lacks a physical cause that explains its exact position, challenging the core assumption of causal determinism from Papineau's point of view. Merely specifying the probabilities with which certain events might occur does not suffice to explain the concrete outcomes that result. 

A common stance among physicalists is to treat the probabilistic nature of quantum mechanics as equivalent to randomness. This view disregards the possibility that quantum events could have a source beyond the physical realm, even while conforming to the probabilities predicted by quantum theory. Unless this non-physical possibility is definitively ruled out, the appeal to quantum mechanics to support causal closure becomes circular: the claim that quantum effects are causally closed relies on the assumption that individual outcomes are random, which in turn is justified only if one already accepts physicalism. Therefore, in the quantum context, the kind of causal closure needed to support physicalism must presuppose it by assuming that randomness itself is a sufficient explanation of the outcomes \cite{persons}. As we have shown, the openness of quantum mechanics allows us to include protophenomenal aspects as an intrinsic quality associated with quantum events and states. In the case of many animals, the existence of causally effective phenomenal aspects explains how perceptions are structured and progressively acquire informative capacity thanks to Darwinian evolution, and at least in the case of humans, allows to include freedom as a determining element of personal identity. These phenomenal properties are consistent with the mathematical description provided by quantum mechanics, but not reducible to it.

Nietszche’s nihilistic turn was the product of an effort to incorporate the scientific view of his epoch as the fundamental basis of his thinking. He considered the mechanistic worldview as a guiding principle that required the greatest discipline and leaving all sentimentality aside. A mechanistic world and its corresponding causal closure leads to question why we should take morality seriously if the world is basically amoral.In other aphorism of The Gay Science \cite{(1882)}  he adds p.346: "We have been boiled down and become cold and hard in the insight that the world proceeds in a manner that is not at all divine, that even by human standards it is by no means rational, merciful, or just: we know that the world in which we live is ungodly, immoral, 'inhuman'." 

Ultimately, the nihilist affirms consciousness above and despite everything \cite{nishitani}. It is the ideal of Nietzsche's Übermensch, who affirms every moment of life despite the pain it may entail, and of Dostoevsky's underground man. For him \cite{underground} Part 1, Chapter IX, “Consciousness is the greatest misfortune of man, yet I know man loves it and will not give it up for any satisfaction. Consciousness, for instance, is infinitely superior two times two makes four ---[to reason]”. But what is the point of consciousness in a mechanistic world? What is the point of this suffering consciousness that the Universe has created, if it is destined to be the consciousness of a world that "is ungodly, immoral, 'inhuman'"?

\subsection{Existentialism}

A particular consequence of the quantum panprotopsychism presented in our previous work, where self-conscious individuals are free due to the effectiveness of consciousness in a quantum world \cite{c3}, is in line with one of the main assumptions of existentialism. The notion of freedom is closely linked to self-conscious persons capable of conceptually and symbolically analyzing their environment and reflecting on future actions and objectives, evaluating risks and possibilities. Our daily personal experience suggests that we are contingently brought into existence. Human beings are driven by both needs and desires. From an early age, we learn through harsh experience the realities of limitation and powerlessness. “The fact that we have desires that we cannot afford to fulfill, and talents and interests that we cannot develop with the time, energy, and resources at our disposal, means that life is a constant succession of choices... we must continually compare and contrast the merits of discordant objectives and prioritize our desiderata" \cite{rescher}. This characterization of persons and the basic features of their personal identity imply that they are essentially incomplete beings. 

Existentialists attempt to idealize this condition by considering that beyond the physical constraints imposed by nature and environment, the agent’s radical freedom knows no bounds. It is entirely responsible for shaping itself: its existence is an ongoing questioning. One’s being in the world is defined not by what one is, but by what one could become: it is a project in perpetual formation. Constantly seeking a sense or place, yet unable to fully define the possibilities of existence, one often experiences life as directionless and devoid of inherent meaning. Rather than simply existing in the world, we find ourselves ``thrown into the world” \cite{Heidegger}. As self-conscious individuals, we are always unfinished projects, navigating existence without a predetermined course. Often, we slip into daily routines that make us forget this fundamental condition.  According to Kierkegaard, angst is a liberating sensation that exposes the emptiness of everyday life. He argues that we experience anxiety or dread precisely because of our radical freedom: the ability to choose without external constraints. For Kierkegaard, what deserves to be thought of is our condition of the single, finite, responsible, simple, suffering, and guilty creature, who has to make a decision in the face of God and who consequently is more interested in ethical questions and in salvation than in abstract speculations \cite{rempel}. 

The development of existentialism as an intellectual movement coincided with the rise of nihilism in 19th-century Europe, sharing a critical view of the Hegelian system, considering that it makes the subject accidental, and thereby transforms existence into something irrelevant. Existentialism is an attempt to embrace our finite condition as definitive and find some way to live with it without denying the negative aspects mentioned. For example, Kierkegaard says in \cite{Either/Or} p.290 that boredom ``can be traced back to the very beginning of the world. The gods were bored; therefore they created human beings. Adam was bored because he was alone; therefore, Eve was created. Since that moment, boredom entered the world and grew in quantity in exact proportion to the growth of population… Boredom is the root of all evil…. Boredom rests upon the nothing that interlaces existence; its dizziness is infinite, like that which comes from looking down into a bottomless abyss.” The search for an escape from boredom generates that need, recognized by Kierkegaard, to find something accidental that provides interest or entertainment. This search characterizes what he calls the aesthetic level of existence.  

For existentialism, the human condition is revealed through an examination of the ways in which we concretely engage with the world in our daily lives and struggle to make sense of and give meaning to our existence. This means that our essence is not given in advance; we are contingently thrown into existence and are burdened with the task of creating ourselves through our choices and actions. we are free and responsible for who we are and what we do. Authenticity is understood to refer to a life lived with a sense of urgency and commitment based on the meaning-giving projects that matter to each of us as individuals. For the existentialist, a moral or praiseworthy life is possible. It is one where we acknowledge and own up to our freedom, take full responsibility for our choices, and act in such a way as to help others realize their freedom \cite{sepexist}. Although existentialism appears to overcome nihilism by affirming an authentic way of life, it is still underpinned by its transient and meaningless condition. For both Kierkegaard and Heidegger, to live authentically, then, is to embrace the uncertainty of existence while continually facing the inevitability of death. As Dostoevsky writes \cite{underground}: “What man wants is simply independent choice, whatever that independence may cost and wherever it may lead. And the choice, of course, the devil only knows what choice” 

Existentialism is also a product of the secularization resulting from scientific mechanisms and the physicalist view.  Having transformed the world in which our lives have meaning [the world of God, morality, will to truth, humanity] into an unbelievable figment and opened up a purposeless and meaningless world as the real world.  Making sense of our lives in this world is the very commitment that is required of us \cite{(1882)}. For this reason, existentialism ended up evolving into a form of atheism. \cite{nishitani} p.185 "According to Sartre, if God existed and had indeed created us, there would be basically no human freedom...There is nothing at all, at the ground of existence. And it is from this ground of "nothing", where there is simply nothing at all, that existence must continually determine itself" Self-creation out of nothing is what Sartre calls freedom \cite{sartre}. In short, existentialism attempts to answer the nihilistic belief that life has no intrinsic meaning or value by asserting that through a combination of free will, awareness, and personal responsibility, we can give meaning to our lives.
Both perspectives share a skepticism towards the existence of objective truths. However, existentialists hold that in matters of existence, the most profound truth one can reach is subjective and that this personal truth is sufficient to reveal the passionate and urgent sense with which we must approach the question of how to live.

\subsection{Anti-foundationalism}

On a philosophical level, nihilism has paved the way for the postmodern response, which manifests itself as either mild annoyance or, more intriguingly, a positive embrace of meaninglessness. For instance, according to Rorty, we no longer have the ability to overcome contingency and pain by appropriation and transformation, but only the ability to recognize them. \cite{rorty}. Unlike Nietzsche’s anxieties and the existentialists' despair, nihilism becomes for anti-foundationalists merely another feature of our modern environment, best approached with calm detachment \cite{nihilism}. 

Anti-foundationalism is a philosophical position that rejects the notion that knowledge must rest on indubitable, self-evident truths or principles. Instead, it views knowledge as a product of social, cultural, and historical contexts that is always subject to revision. Because it denies the existence of rationally grounded, self-evident truths, anti-foundationalism often casts doubt on the possibility of objective, universal truths or overarching narratives, including those offered by metaphysics. It opposes totalizing accounts of social, scientific, or historical reality, viewing them as illegitimate, and favors localized, context-specific narratives instead. However, this stance may seem excessive, especially when it leads to the dismissal of forms of evidence grounded in lived meaningful experiences shared across diverse human traditions. Yet, as anti-foundationalist thinkers like Rorty and Foucault \cite{foucault2} have implicitly acknowledged, their critique aims less at denying universal modes of experience and more at challenging traditional metaphysical or metaethical frameworks solely based on rational or semantic analysis.

It is important to note that many anti-foundationalist positions are merely the natural conclusion of a process that began at the dawn of modernity and has produced seemingly irreversible changes in our understanding of human beings and the world in general, which are not questionable.  We have learned that neither in science nor in philosophy does it seem possible to attain absolute truths. We already mentioned Hume's skepticism that questioned the assumptions of scientific inquiry with the problem of induction. We can only have well-founded beliefs in scientific matters; we cannot be sure that what has happened today will continue to happen tomorrow. But the most practical thing is to assume that the world has an orderly nature and that if we proceed in this way, that is, trying to discover how things behave through experience and reason, we will not be surprised. 

We do not dispute the adoption of a pragmatic approach to the value of scientific or philosophical knowledge, but we do not consider every claim about the nature of the world to be not only unnecessary but untenable, as many anti-foundationalist pragmatists do. Claiming that the most practical course of action is to assume that the behaviors we have observed in the past will continue to be those observed in the future is untenable without assuming the ordered nature of things and ends up casting the problem of induction in different terms: we end up believing it is more practical to proceed this way because it has worked so far. The attempts of antifoundationalists such as Richard Rorty, Daniel Dennett, and David Armstrong to draw a parallel between the mind and a thermostat and to deny our subjective experiences are not only untenable but also fruitless. They are based on Sellars's strategy of denying the existence of entities that cause seemingly insoluble problems. Following philosophers such as Thomas Nagel, John Searle, and David Chalmers, we believe that adopting a scientific position does not entail denying the existence of subjective experiences. 

Anti-foundationalism also presents valuable aspects as the dismissal of an erroneous view about the existence of a rationally recognizable foundation for ethics, as sought by Plato. Anti-foundationalism considers that if there is a source of certainty about our behavior, it does not come solely from rational analysis but because of our way of feeling. Rorty himself states that what makes us different from other animals is not that we can know and they merely feel, but that perhaps we can feel for one another to a much greater extent than they can. He considers that this would allow us to dissociate Christ's proposal that love matters more than knowledge from the Neoplatonic idea that knowledge of the truth will set us free.

But the ultimate expression of anti-foundationalism is its denunciation of what Rorty calls the ``attachment to the idea of universal validity” \cite{rorty2}, which he considers a variety of the temptations that led Plato, Augustine, Kant, Nietzsche, Heidegger, among many others to seek a relationship with something greater than themselves and the contingent circumstances in which they found themselves (for example, the good, God, the moral law, the will to power, being). This is the point where our positions diverge more acutely as we discuss in the next section. 

\section{A consciousness centered world}

If experimental evidence were to uncover quantum phenomena in the brain that enable the emergence of stable entangled states linked to neuronal activity, a form of quantum panprotopsychism would become highly credible. This point of view would be further reinforced by the fact that within such a framework, the combination problem traditionally associated with panpsychism would cease to pose a challenge. At the same time, it is undeniable that biological evolution has progressively given rise to species with increasingly sophisticated mental and conscious capacities. Such a discovery would compel a fundamental rethinking of the prevailing worldview shaped over centuries by classical physics. Within this new perspective, consciousness would not be regarded as a secondary outcome of physical processes, but rather as a fundamental feature of reality. There would be plausible evidence on the phenomenal nature of a world that should be considered as centered in consciousness.

Quantum panprotopsychism could provide a framework for understanding the intrinsic nature of reality by linking the phenomenal properties we access through subjective experience to their physical counterparts. In quantum mechanics, a system’s state is represented by a vector in Hilbert space, encapsulating all the information available about the system and enabling probabilistic predictions of its future behavior. If physical states are mathematical expressions of internal phenomenal states, whose qualitative aspects remain inaccessible to external observers, then their intrinsic nature may extend beyond what their physical description alone can capture.

Quantum indeterminism suggests a conception of the ‘self’ grounded in agents capable of free actions, potentially guided by ethical or aesthetic considerations. Consciousness, in this view, must play an active role in generating increasingly rich, complex, and informative perceptions. If physical and mathematical laws merely constrain a phenomenal reality that may transcend them, as the causal openness of quantum mechanics seems to suggest, what is the nature of that reality, and how might humans, as the most advanced products of conscious biological evolution, gain access to it? What new understanding arises from the recognition that quantum mechanics, the most fundamental of our physical theories, points toward an intrinsically phenomenal reality? Given the causal openness of quantum theory, perceptions and feelings may actively influence outcomes, as previously proposed, by enhancing the adaptive capacities of organisms, increasing informational value, and contributing to evolutionary progress. This opens the possibility that purely phenomenal elements exist, those that shape the structure of consciousness or guide our free decisions.

Perhaps the question of where the development of consciousness ultimately leads is intimately tied to the question of its very existence, a question that has preoccupied many philosophers, including Leibniz and Heidegger, who regarded it as the fundamental question of metaphysics: Why is there something rather than nothing? Heidegger formulates this inquiry as follows \cite{Heidegger5}, p.3: 'Why are there beings at all?... From what ground do beings come? On what ground do beings stand? To what ground do the beings go…?' He further notes that it remains unresolved whether this ground offers any foundation at all. The question challenges the status of what exists, the entities that science investigates. Yet, according to Heidegger, metaphysics inquires about beings while neglecting the being of those beings. For metaphysics, being has dissolved into an illusion, as Nietzsche suggests. In a universe centered on consciousness, however, this question shifts toward an investigation of human beings as the highest expression of conscious awareness. Within such a framework, it becomes plausible that the question of why anything exists at all might be answerable if consciousness itself lies at the heart of reality. And perhaps the existential need to question the meaning of being would vanish once individuals attain the capacity to act with love and creativity, while experiencing a profound sense of fulfillment. But then what could be the nature of this center?

Heidegger sought the answer about the being in the intuitions of the pre-Socratics, particularly Parmenides. In his poem, the goddess of the night begins by stating, “’What Is’ is ungenerated and deathless, / whole and uniform, and still and perfect” \cite{sepparm} manifests itself in an eternal present: “but not ever was it, nor yet will it be, since it is now together entire,/ single, continuous; for what birth will you seek of it?/”. It encompasses the entirety of what exists “Nor is it divided, since it is all alike;/ and it is not any more there, which would keep it from holding together, / nor any worse, but it is all replete with ‘What Is’./ Therefore it is all continuous: for ‘What Is’ draws to ‘What Is’.” Parmenides' ‘What Is’ could then be considered as the center to which the evolution of consciousness points. John Palmer says \cite{sepparm} about the ‘What Is’: “It thus seems preferable to understand ‘What Is' as coterminous but not consubstantial with the perceptible cosmos: On this view, ‘What Is’ imperceptibly interpenetrates or runs through all things while yet maintaining its own identity distinct from theirs. ...Parmenides’ vision of the relation between ‘What Is’ and the developed cosmos, as coterminous but not consubstantial, also has its analogue in Xenophanes’ conception of the relation between his one greatest god and the cosmos, as well as in Empedocles’ conception of the divinity that is the persistent aspect of the cosmos’ perfectly unified condition, darting throughout the cosmos with its swift thought. Both appear to be coterminous but not consubstantial with the cosmos they penetrate.”
It is not accessible to ordinary knowledge, and accessing it requires a special spiritual disposition. In a panpsychist framework, "What is" is phenomenal and has a perfectly unified condition, lacking any physico-mathematical description. A structureless, phenomenal  no-thingness \cite{Nishida}  ‘whole and uniform and still and perfect’ that could be the source of our own ethical and aesthetical values. Phenomenal states that Richard Jones and Jerome Gellman \cite{SEPMysticism} characterize as follows: ``A putatively non-sensory consciousness or unstructured sensory experience that grants knowledge of realities or states of affairs that are of a kind not accessible through ordinary sensory perception structured by mental conceptions, somato-sensory modalities, or standard introspection."

Many reports of this kind of experiences appear throughout history in many contexts and religions. For instance, “My mind,” says St. Augustine in his Confessions \cite{Augustine} Ch.xvii  “withdrew its thoughts from experience, extracting itself from the contradictory throng of sensuous images, that it might find out what that light was wherein it was bathed... And thus, with the flash of one hurried glance, it attained to the vision of That Which Is ..., but I could not sustain my gaze: my weakness was dashed back, and I was relegated to my ordinary experience, bearing with me only a loving memory, and as it were the fragrance of those desirable meats on the which as yet I was not able to feed.”

Some of our everyday conscious experiences are not communicable to others who, due to limitations—for instance, in their sensory organs—lack access to similar experiences. Their existence cannot be verified by those who have never undergone them. Even among conscious beings, comparisons between their experiential states are merely indexical --—for example, by pointing to an object of a particular color—-- allowing us to agree that our respective experiences refer to the same object, but not to determine whether the qualitative aspects of those experiences are the same. Mystical experiences, too, are phenomenal in nature, presenting themselves as conscious contents, yet they lack any ostensible or publicly shareable dimension. It is often said that such experiences reveal more than can ever be articulated in language, without denying, however, that something meaningful can still be expressed about them, as was the case with Parmenides\cite{SEPMysticism}.

It would constitute a unifying experience of the universe, grounded in a reality described differently across civilizations, a 'no-thing' \cite{Nishida} that is shared by all beings. Such an experience would fundamentally affirm the intrinsic value of every conscious personal life, independent of space and time. It would transcend all conceptual frameworks and remain accessible to individuals of a wide range of cultural and spiritual traditions.  Although this type of experience has occurred in the most diverse societies, it is extremely rare. Its exceptional nature points to the need for a radical change in the basic functioning of self-aware beings. It requires abandoning our habitual way of thinking. An idea of detachment or liberation seems to be the common denominator of every approach to this exceptional state. It requires a profound change in our mental functioning and freeing ourselves from deeply rooted forms of behavior. Eckhart says in his sermons about the Eternal birth: “None assuredly can experience or approach this birth without a mighty effort. A man cannot attain to this birth except by withdrawing his senses from all things. And that requires a mighty effort to drive back the powers of the soul and inhibit their functioning. This must be done with force; without force it cannot be done. As Christ said, “The kingdom of heaven suffers violence, and the violent take it by force” (Matt. 11:12 ). Ian Alexander Moore in his book {\em Eckhart, Heidegger and the Imperative of Releasement} \cite{moore} adds based on notes by Heidegger entitled 'irrationality in Meister Eckardt'.: “what we need is a detachment that is not just ‘negative’ but has a ‘positive side, one that is at work even as we direct ourselves to the world. This positive unification through detachment would primarily not be of a theoretical nature, but rather of a lived, ‘emotional’, ‘religious’ experience. There is thus something ‘irrational’ about this central concept: ‘detachment’. Heidegger adds that this concept for Eckhart implies the suspension of ‘the understanding, as judging, pulling apart into the duality of subject and predicate.’  This is achievable for us if the ‘self’ does not distance itself from things and set itself up as a judge who measures, judges, and evaluates everything, establishing an illusory distance from the whole of which it is a part \cite{isava} p. 58. 

These conscious states appear to infuse the phenomenal experience with a distinct affective tone. But what is the origin of this tone? And is there a physical counterpart, if any, that accounts for its effectiveness and the behaviors that follow from such experiences? Knowledge of the conscious state alone does not determine behavior, as quantum mechanics only allows probabilistic predictions, not precise outcomes. For this affective tone to remain consistent with physical theory, it is sufficient that it does not alter the system’s range of possible behaviors (i.e., the person's behavioral repertoire), but merely adjusts the conditional probabilities—just as any set of choices would. The emergence of a purely phenomenal dimension, a kind of nothingness, is thus fully compatible with the quantum physical framework. Although it may profoundly transform the internal states of a person, it only becomes manifest in the decisions-making processes that precede quantum measurements. Free acts, such as the creation of an artistic masterpiece, an expression of generosity, compassion, heroism, or even the experience of remorse, would be grounded in this very nature.

But what kind of detachment or release is required. As we have already analyzed, to reach forms of phenomenal manifestation richer than simple choices at the end of prehensions, the What Is, it is necessary to interrupt the conceptual processes that intervene when we strive for the development of the 'me'. The constant desire to be something in this world and something that can endure beyond death must end. But the root of the self and its desire for indefinite continuity lies in the agent who permanently chooses to continue existing. That is what Nietzsche calls the spirit of revenge. As Nishitani observes \cite{nishitani} p.1: If nihilism is anything, it is first of all a problem of the self. And it becomes such a problem only when the self becomes a problem, when the ground of existence called ‘self’ becomes a problem, for itself. When the problem of nihilism is posed apart from the self or as a problem of society in general, it loses the special genuineness that distinguishes it from other problems ...This is what makes the question of nihilism the radical question it is.”

We are troubled by the reality of death and seek within our 'self', in our 'me', something of lasting value: an act, a love, a creation. Yet, it is only through a kind of inner dying, a relinquishment of the self while still alive that the truly new can emerge. Nishitani describes this as the transcendence of the division between the observing self, which we might identify with the 'I', and the observed object. He writes (p.2): 'Here subjectivity in the true sense appears for the first time: the standpoint arises in which one strives resolutely to be oneself and to seek the ground of one's actual existence. It is also here that nihility is revealed for the first time. By being thrown to nihility, the self is revealed for the first time.' Through this encounter, the being of the self is unveiled to itself as groundless. In such a process of contemplation and awareness, the foundation of existence itself is disclosed. 'And because God is no-thing [i.e., not subject to physical laws], there is no place where God is not' \cite{Nishida}.

Putting an end to the cultivation of the ‘me’ is a first step toward overcoming the dichotomy of the observer and the observed. Among other behaviors oriented by this cultivation, we can recognize the following: i) Actions with an end in view, with the desire for a result, the pursuit of an idea, publishing a book, earning money, winning in a sport. The idea and the action that pursues it are necessary but not sufficient conditions for the egocentric action involved in the adoption of an objective. It is the constant effort to control our 'me' to be something different from what one is or keep an image that we consider desirable. ii) Identification with a profession, religion, or any intellectual or physical activity is nothing but another way of strengthening the 'me'. I am the one who knows how the world really is, I am above fear and doubt, I know that God exists, and I have an immortal soul. All of these behaviors hide the same finality of strengthening the ‘me’. iii) Self-deception: believing that we are accomplished or recognized when we manage to convince others not only of a certain ability but also of the kind of people we are, good, brave, wise, or whatever. When we manage to convey a certain way of conceiving the self as interesting, attractive, or admirable. This form of self-deception through the search for recognition has proven to be one of the main driving forces of history \cite{Kojeve}.

The ideal of monastic life, which began to take shape in the West during the fourth and fifth centuries \cite{agamben}, was conceived as a means of neutralizing the influence of the 'me'. Initially envisioned as a solitary and individual retreat from the world, it gradually evolved into a communal model that meticulously regulated every aspect of a monk’s existence. Solitude not only removes one from the joys of shared celebration and the sorrows of others’ suffering but also from the self-awareness that arises through human relationships and the recognition of one's faults within them. The coenobium, a communal form of monastic life, imposes a habitus: a disciplined mode of dressing, dwelling, and living that shapes every moment of the monk’s day. The monastic habit, as clothing, is intended to suppress any form of individual distinction, forbidding attire that lacks meaning or virtue. Every hour of the monk’s day is regulated, from duties to prayers, requiring continuous attention to the task at hand. In this way, any desire stemming from the ‘me’ that does not align with the communal and devotional pursuit of God’s will is systematically subdued. To this, Francis of Assisi added a radical dimension: the renunciation of all property rights—the refusal to claim ownership of anything as one’s own. In monastic life, any form of cultivation of the 'me': actions with a personal objective, identification with certain personal qualities or goods, self-deception, and the search for recognition are strongly combated.

Developing our capacities to feel and perceive requires freeing ourselves from a self-centered life. To achieve this, we need a behavior that neither approves nor condemns actions, thoughts, or feelings, and the elimination of all distractions that lead us to avoid, evade, or refuse to pay attention to the present moment. This not only requires a high degree of sensitivity, but also a relinquishment of the pursuit of personal goals. This way of being involves being simple, paying attention to the present, and being receptive to what happens to us in each moment without seeking anything or pursuing any goal. Having the humility to take a step back. The effectiveness of consciousness is essential for this change, and it is in consciousness itself that change occurs. Realizing that we are failing in our attempt to free ourselves from our 'me' may be relatively simple, but maintaining constant attention on that failure is extremely difficult and constitutes the essence of the sacrifice that those who follow this path claim is required. The necessary state of attention, if transformed into effort, produces frustration or reinforces the self; and even when this does not happen, it requires avoiding all forms of distraction or entertainment, as well as the disappointments and meaninglessness the self feels at its repeated failures to achieve this detachment.

When the 'me' centered activity ceases one can reach a state that Kierkegaard characterizes as anxiety where one is confronted with the  \cite{kierkegaard} p.20 “dizziness of freedom on the brink of the abyss, where the self, ...looks into the abyss as the infinite possibility within itself.” When it chooses to cling to its finitude putting an end to his freedom, it opts for selfishness, and the original sin buried in the depths of freedom comes to the surface \cite{Kierkegaard6}. For Kierkegaard this confrontation with the original sin and the finitude is at the same time the opportunity to overcome its condition. That would affect our entire phenomenal structure: the way we perceive and value the world and our relationship with it. It is a change that can occur in anyone and can be applied to our way of relating to the world and to people, awakening a loving and reverent bond. The change does not occur in what is observed, but in the way the subjects observe it. One is not just aware of something; it is part of something. It is simply something greater, eternal, and completely self-sufficient. Time is no longer perceived as the inevitable becoming, rejected by Nietzsche. Each moment, with all its vitality, has an absolute and lasting value. Transcending is not abandoning this world to reach another that is outside of sensible reality, but rather going to meet the numinous within this world. \cite{schajowicz}

Consciousness is effective, as suggested by quantum panprotopsychism. Agents can, without violating the limits of quantum description, express themselves in free choices originating from a purely phenomenal realm, physically limited but not supervening on the objects described by the physical laws. We inhabit a consciousness-centered world that through a rigorous physical order leads to the gates of a transcendent reality, one that certain individuals can manifest within the immanent realm. With the liberation of the 'me', one can become vulnerable and highly sensitive to the world. It is about understanding this constant movement that sustains the 'me' with direct and instantaneous discernment. When this occurs, one reaches what Nishida calls 'true knowledge', where \cite{Nishida} p.111 ''the presence of conscious phenomena and our consciousness of them are straightforwardly one and the same, with no interval for subject to be distinguished from object." Both subject and object occur in consciousness at the same level. Quantum panprotopsychism could be a step towards the understanding of the relation between reason and 'true knowledge' observed by the thinkers of the Kyoto School of Philosophy, in particular, in the field of What Is, things do not exist as objects of knowledge, since rationality ---the subject’s ability to represent a thing as ‘object’ --- is no longer operative. In this field knowledge is free of the bifurcating ego, it is a knowledge in which contact with the reality of a thing is at the same time contact with the reality of the self.  \cite{horo}.

What is discussed here does not refer to another world. Quantum panprotopsychism sets aside any dualism and refers to a purely phenomenal world governed by quantum laws. Mystical experiences, although part of the most diverse religions have become independent of them, and refer to the most intimate nature of the world in which we live and of which we are a part.

Sloterdijk \cite{Sloterdijk} p.186 summarizes this point of view about the meaning of a contemporary mysticism as follows: "the world is an arena of ineluctable passions...To exist is to play a role ---the 'role' being that of subjects capable of conflict ...who behaves heroically, ...in their unavoidable struggles for chances in life. ... If someone were to come up with the view that the map is not the land...[our] self-image is not oneself... that would be total news to them. ...Exactly this would be the mystical message... Mystics are the contra-heroic informants on the human being; ...They try to rebut the arena ontology as such. They do so by showing that the something in which we 'reside' is in truth an unmarked space in which there can be no difference...that justifies struggles for life and death along the line of fighting for the position of one's ego."

\section{Conclusions}

In a series of papers, we have shown that a form of panprotopsychism rooted in quantum states and events can bypass the combination problem. The entangled states in quantum mechanics provide a type of emergence, where new properties of the whole arise that do not simply supervene on the properties of the parts. This accounts for how complex phenomenal qualities and subjects emerge. Quantum mechanics is causally open, allowing new properties to materialize during measurement processes due  to its inherent indeterminacy, without physical causes. This openness enables consciousness, and more broadly, phenomenal properties, to have causal efficacy. It would enhance the adaptive capacities of animals, while also explaining the intricate structure of our perceptions.

Persons are self-conscious beings capable of conceptually and symbolically analyzing their environment and reflecting on their future actions and objectives, evaluating risks and possibilities. Life is a constant succession of choices... we must continually compare the merits of discordant objectives. A person’s characterization involves a physical substratum, her brain, her perceptions and actions. Thus, persons are agents. For agency, a robust first-person perspective intimately linked to its capacities for symbolic expression, language, and conceptualization is needed. We act with the desire for a result, the pursuit of an idea, publishing a book, earning money, winning in a sport. 

We often act for the strengthening of the 'me' which is a conceptual construction. As science advanced and the world came to be understood as a mechanistic system governed by classical laws, various intellectual movements ---ranging from nihilism to existentialism--- emerged in late nineteenth-century Europe. This shift coincided with the decline of the pre-modern religious worldview, giving way to an increasingly secular and scientific perspective. As a result, the loss of a transcendent moral framework contributed to the defining experiences of modernity: anxiety, alienation, boredom, and meaninglessness. If the vision resulting from quantum panprotopsychism is correct and can be confirmed experimentally by observing entangled quantum states coupled to our neural system, we live in a consciousness-centered world. This would allow us to conceive the existence of a purely phenomenal manifestation within us that does not necessarily have a completely physical counterpart that is not fully accessible with a ‘me’ centered life. Thus, a world centered on consciousness can lead, through a rigorous physical order, to the evolution of personal beings capable of accessing a transcendent phenomenal realm that can manifest itself in the immanent domain. The overcoming of the 'me'-centered activity would lead to a form of consciousness that excludes nothing and puts subject and object at the same level. ``When a man sees All in all, then a man stands beyond mere understanding" \cite{Eckhart}.

In summary, quantum panprotopsychism invites us to conceive of a universe fundamentally centered on consciousness. According to this view, the states and events described by the formalism of quantum mechanics correspond --—following van Fraassen’s interpretation—-- to brain processes that possess a phenomenal counterpart. The physical openness of quantum mechanics, grounded in its indeterminism, implies that some physical effects may arise without physical causes, instead originating from purely phenomenal sources. As a result, the phenomenal domain transcends the boundaries of physics: phenomenal contents are mirrored by quantum states only insofar as they represent causally regulated processes, while their source lies in a transcendent sphere of action beyond physical determination. It is within this sphere that the pure reception of the Parmenidean 'What Is' could take place. Such experiences, being purely phenomenal, unlike events, manifest themselves as conscious contents that lack any ostensible or publicly accessible expression but that, according to quantum mechanics, would be capable of changing our behavior.

\section{Acknowledgements}
I wish to thank Jorge Pullin for help with preparation with this manuscript. This work was supported in part by grant  NSF-PHY-2206557, funds of the Hearne Institute for Theoretical Physics, CCT-LSU, Fondo Clemente Estable FCE 1 2023 1 155865.


\begin{thebibliography}{99}
\bibitem{dennett1}
Dennett, D. (1992) ``Consciousness explained", Back Bay Books, New York, NY.

\bibitem[Gambini and Pullin(2024a)]{c1}
 Gambini, R. and Pullin, J. (2024)
`Quantum panprotopsychism and the combination problem,''
Mind and Matter, 22 (1). pp. 51-94.

\bibitem{husserl}
Woodruff Smith, David (2006) ``Husserl'', Routledge, New York.

\bibitem{low}
Low, P. (2012)``The Cambridge Declaration on Consciousness'' https://philiplow.foundation/consciousness/

\bibitem{gija} Ginsburg, S. and Jablonka, E., ``The evolution of the sensitive soul. Learning and the origins of consciousness'', MIT Press, Boston, MA, USA (2019).

\bibitem
{c2}
 Gambini, R. and Pullin, J. (2024)
``Quantum panprotopsychism and the structure and subject-summing combination problem,''
[arXiv:2409.01368 [q-bio.NC]], to appear in J. Consc. Stud.\

\bibitem{SEPMysticism}
Jones, Richard and Jerome Gellman, "Mysticism", The Stanford Encyclopedia of Philosophy (Summer 2025 Edition), Edward N. Zalta \& Uri Nodelman (eds.), 

\bibitem{Feshbach1975} Feshbach, N. (1975) "Empathy in children", The counseling Psychologist, 5, 25.

\bibitem{prvdw} Preston, S., De Waal, F. (2002) "Altruism and altruistic love science, philosophy and religion in dialogue, Oxford Academic, Oxford, UK.

\bibitem{baker2} Baker, L. (2003) "The Difference that Self-Consciousness Makes" in Petrus, K. "Onhuman persons", Ontus, NewYork, NY.

\bibitem{Kierkegaard6} Kierkegaard, S. (2014) "The concept of anxiety", Liveright, New York, NY.

\bibitem
{seagersep}
Goff, P., Seager, W., and Sean Allen-Hermanson, "Panpsychism", The Stanford Encyclopedia of Philosophy (Summer 2022 Edition), Edward N. Zalta (ed.), URL = https://plato.stanford.edu/archives/sum2022/entries/panpsychism/.

\bibitem
{lulegues} Gambini, R., Lewowicz, L., Pullin, J. (2015) {\em Found. Chem.} 17, 117-127.

 \bibitem{healey}  Healey, R, (1999) 
``Holism  and nonseparability in physics'' in Stanford 
Encyclopedia of Philosophy 

http://plato.stanford.edu.


 \bibitem
 {teller} P. Teller,  (1986) {\em Brit. J. Phil. Sci.} 37, 71.

\bibitem{chalmerscp}
Chalmers, D. (2016) 
``The Combination Problem of Panpsychism'' in Panpsychism, Contemporary Perspectives. Godehard Bruntrup (ed.), Ludwig Jaskolla (ed.), Oxford University Press, UK.

 \bibitem
 {davies3} Davies, P. (2014) ``Information and the Nature of Reality: From Physics To Metaphysics", Cambridge University Press, Cambridge, UK.

 \bibitem
 {vfr2} van Fraassen, B. (2008)  "Scientific Representation: Paradoxes of Perspective" Clarendon Press - Oxford, NY.

\bibitem{c3} Gambini, R. and Pullin, J. (2025) ``Self-consciousness and personal identity in quantum panprotopsychism".

\bibitem{korfmacher} Korfmacher, C. (2006) ``Personal identity", Internet Encyclopedia of Phyilosophy.

\bibitem
{Davies3} Davies, P. (2014) ``Information and the Nature of Reality: From Physics To Metaphysics", Cambridge University Press, Cambridge, UK.

 \bibitem
 [Fisher(2015)]
 {fisher} Fisher, M. (2015), {\it Annals of Physics}, 362, 593.

\bibitem{ep1} Fuchs, C. A., Mermin, N. D., Schack, R. (2014).
"An introduction to QBism with an application to the locality of quantum mechanics."
American Journal of Physics, 82(8), 749–754

\bibitem{ep2} Rovelli, C. (1996).
"Relational quantum mechanics."
International Journal of Theoretical Physics, 35(8), 1637–1678.

\bibitem{leifer} Leifer, M. S. (2014).
"Is the quantum state real? An extended review of $\psi$ ontology theorems."
Quanta, 3(1), 67–155.

\bibitem{whiteheadaoi} Whitehead, A. (1933) ``Adventures of Ideas", The Free Press, London, UK.

\bibitem{WhiteheadProcess} Whitehead, A. N. (1978) ``Process and reality", MacMillan, London, UK.

\bibitem{peresspri}
Peres, A. (2006) Quantum theory: concepts and methods, Vol. 57
Springer Science \& Business Media, New York, NY.

\bibitem{crisis2} Husserl, E. (1970) "The Crisis of European Sciences and Transcendental Phenomenology: An Introduction to Phenomenological Philosophy (Northwestern University Studies in Phenomenology \& Existential Philosophy)" Northwestern University Press, Evanston, IL.

\bibitem{Korsgaard}
Korsgaard, K. (1996), ”The sources of normativity”, Cambridge University Press, Cambridge, UK


\bibitem{foucault2} (2007) Gary Gutting ed., The Cambridge Companion to Foucault UK.  

 \bibitem{chalmers} 
Chalmers, D. (1996)
``The Conscious Mind: In Search of a Fundamental Theory (Philosophy of Mind)", Oxford University Press, Oxford, UK.

\bibitem
{lrb} Rudder Baker, L. (2013) ``Naturalism and the First-Person Perspective'', Oxford University Press, Oxford, UK.

\bibitem[Wittgenstein(1918)]{wittgenstein}
  Wittgenstein, E. ``Tractatus logico-philosophicus'' Cosimo Classics, New York, NY (2007) original (1918).

\bibitem{choifer}
Choifer, A. (2018) Philos. Papers 47, 333.  

\bibitem{Gazzaniga}
Gazzaniga, M. (2011) ``Who's in Charge?: The Neuroscience of Decision-Making, the Notion of Free Will and the Idea of a Determined World''

\bibitem{Darwin} Darwin, C. (2010) ``The Descent of Man'',Dover, Mineola, NY, USA.

\bibitem
  {dieguez}
  Dieguez, A. (2011)   ``Conceptual Thinking in Animals. Some Reflections on Language, Concepts, and Mind''
  In ``Darwin's Evolving Legacy'' (pp.383-395) Editors: Martínez-Contrera, J. and  Ponce de León, A., Publisher: Siglo XXI y Universidad Veracruzana.


 \bibitem
 {jonas} Jonas, H. (2001) ``The phenomenon of life'', Northwestern University Press, Evanston, IL, USA.

 
\bibitem
{cassirer2} Cassirer, E. (2020) "The philosophy of symbolic forms", Routledge, New York, NY.

\bibitem{SEPenl} Bristow, W. (2017) "Enlightenment" Stanford Encyclopedia of Philosophy

https://plato.stanford.edu/entries/enlightenment/

\bibitem{ttp} Spinoza, B. (1989) "Tractatus Theologico-Politicus"  (Gebhardt Edition 1925) Trans. S. Shirley, Leiden Netherlands. 

\bibitem{Israel} Israel, J. (2001) "Radical Enlightenment" Oxford University Press UK.

\bibitem
{alexander} Alexander, S. ``Space, time and deity'', Wentworth Press, Sydney, Australia (2019) original from (1920).

\bibitem
{foucault} Foucault, M. (1998) ``Ethics: subjectivity and truth'' edited by Rabinow, P., The New Press, New York.

\bibitem
{persons}
 Gambini, R. and Pullin, J. (2022)
``Atoms and persons'', World Scientific, Singapore (2022).

\bibitem
  {tara}
    Tagini, A. and Raffone, A. (2010) Cognitive Processes, 11,9.

\bibitem
{epstein} Epstein, R., (1981) The Behavior Analyst, 4, 42.

  \bibitem
  {pickering} Pickering, (1999) J. Journal of Consciousness Studies, 6, 31.

\bibitem{mcmillan}
McMillan, F. (2005)``Mental Health and Well-Being in Animals" Wiley-Blackwell, New York, NY.

\bibitem{Augustine} Augustine, St. (2024) "Confessions", Ascension, New York, NY. 

\bibitem{Heidegger5} Heidedgger, M. (2000), "Introduction to metaphysics", Yale University Press, New Haven CT.

\bibitem{sepparm} Palmer, J. (2025) "Parmenides" The Stanford Encyclopedia of Philosophy (Spring 2025 Edition), Edward N. Zalta \& Uri Nodelman (eds.). https://plato.stanford.edu/entries/parmenides/

\bibitem{rescher} Rescher, N. (1998) "Complexity: a philosophical overview" Transaction Pulishers, New Brunswick, NJ.

\bibitem{Heidegger}  Heidegger, T. (1927/2008) ``Being and time", Harper Perennial MOdern Thought, New York. 

\bibitem{papineau} Papineau, David, ``The rise of physicalism'' in ``Physicalism and its discontents'', Gillett, C and Loewer, B. (editors) Cambridge University Press, Cambridge, UK (2009).

\bibitem{Either/Or} Kierkegaard, Soren, (1843/1992) ``Either/Or A fragment of life" Penguin Classics.

\bibitem{sepexist} Aho, Kevin, (2025) "Existentialism", The Stanford Encyclopedia of Philosophy (Spring 2025 Edition), Edward N. Zalta \& Uri Nodelman (eds.).

\bibitem{rorty}
Rorty, R. (1989) ``Contingency, Irony, and Solidarity" Cambridge University Press, Cambridge, UK.

\bibitem{rorty2}
Rorty, R. (1995) "Habermas Derrida and the Functions of Philosophy" in
Revue Internationale de Philosophie Vol. 49, No. 194(4) pp. 437-459 


\bibitem{kaufmann} Kaufmann, W. (1961) "Critique of Religion and Philosophy" Doubleday, New York, NY.

\bibitem{moore} Moore, I.A., (2019) "Eckhart, Heidegger and the Imperative of Releasement" State University of New York Press 2019, Albany, NY.

\bibitem{Nishida} Nishida, C. (1990) "An inquiry into the good", Yale University Press, New Haven, CT.

\bibitem{isava} Isava, L.M. (1990) 'Voz de Amante (Estudio sobre la poesia de Rafael Cadenas)" Biblioteca de la Academia Nacional de Historia Vol 136, Caracas 

\bibitem{literature and life} Cadenas, R. (1970) "Literatura y vida" in Letras Nuevas N.6, Caracas 

\bibitem{bollnow} Bollnow, O.F. (1963) "Rilke" Taurus, Madrid 

\bibitem{Heidegger2} Heidegger, M. (1977) "Question Concerning Technology, and Other Essays" Harper Torchbooks, New York, NY.

\bibitem{(1882)} Nietzsche, F. (1882/1974) "The gay science" Vintage Random House, New YOrk, NY.

\bibitem{underground} Dostoevsky, F. "Notes from underground The New Translation by Richard Pevear and Larissa Volokhonsky" Kindle Edition

\bibitem{nickbea} Bea, Nick, https://thekyotoschoolofphilosophy.wordpress.com/the-death-of-god/

\bibitem{Bauman} Bauman, Z. (2000) ``Liquid modernity", Polity, New York, NY.

\bibitem{nishitani} Nishitani, K. (1990) ``The self-overcoming of nihilism" SUNY Press, New York.

\bibitem{sartre} Sartre, J.P. (1943/2018) ``Being and Nothingness, an essay in phenomenological ontology" Washington Square Press, New York 

\bibitem{nihilism} Pratt, A. "Nihilism", Internet Encyclopedia of Philosophy. iep.utm.edu

\bibitem{thinking} Heidegger, M. (1976) "What is called thinking?", Harper Perennial, New York, NY.

\bibitem{Kojeve} Kojeve, A. (1980) ``Introduction to the reading of Hegel", Cornell Universirty Press, Ithaca, NY.

\bibitem{agamben} Agamben, G. (2023) "The Highest Poverty: Monastic Rules and Form-of-Life" Stanford University Press. 

\bibitem{rempel} Rempel, G. (1959) "Soren Kierkegaard and Existentialism" https://www.sorenkierkegaard.nl

\bibitem{kierkegaardeither} Kierkegaard, S. (1992) "Either/or", Penguin Classics, New York, NY.

\bibitem{kierkegaard} Stokes, P. (2015) ”The naked self: Kierkegaard and personal identity”, Oxford Academic, Oxford, UK.

\bibitem{schajowicz} Schajowicz, L (1990) "Mito y Existencia" Editorial de la Universidad de Puerto Rico, PR.  

\bibitem{horo} Horo, A. (1992) "A Call for a Paradigm Shift in Philosophical Thought" Nanzan Bulletin 16, 1992

\bibitem{Sloterdijk} Sloterdijk, P. (2020) "After God" Polity First edition, UK.

\bibitem{Eckhart} Eckhart, M. (1999) "The complete mystical works" Crossroads, New York, NY.

\bibitem{machado} Machado, A. (1971) "Nuevas canciones. De un cancionero apócrifo. Edición de José María Valverde.' Ed. Castalia, Madrid.
\end{thebibliography}
\end{document}